\documentclass{article}

\input{tcilatex}
\begin{document}

\centerline{The linear relations between the complex moduli and using the
linear relations}
\centerline{reduce the stabilization equations for
supersymmetric black holes in N=2 theory}

\centerline{Dafa Li} \centerline{Dept of mathematical sciences} %
\centerline{Tsinghua University, Beijing 100084, China}
\centerline{emails:
lidafa@tsinghua.edu.cn}

Abstract: In [Phys. Rev. D 54, 6293 (1996)], the black hole entropy was
derived by solving the matrix equation obtained from the stabilization
equations for the solution of frozen moduli. In this paper, by directly
solving the stabilization equations for the solution of frozen moduli
without using the matrix equation, we find linear relations between any two
of the three complex moduli at the black hole horizon. So far, no one
discusses the linear relations. Via the linear relations, we derive the
unique solution of frozen moduli and reduce the stabilization equations in
three different ways. For example, the eight stabilization equations in
[Phys. Rev. D 54, 6293 (1996)]\ can be replaced equivalently with three
equations: the solution of moduli $z_{1}$ and two linear relations, which
are much simpler and more intuitive.

\section{Introduction}

The universal entropy-area formula for supersymmetric black holes in $%
N=2,4,8 $ theories was derived \cite{Ferrara-1, Ferrara-2}. The general form
of the central charge for the $N=2$ theory is given in \cite{Ceresole,
Behrndt, Behrndt-2, Kallosh96}
\begin{equation}
Z=e^{K/2}(X^{\Lambda }q_{\Lambda }-F_{\Lambda }p^{\Lambda })=L^{\Lambda
}q_{\Lambda }-M_{\Lambda }P^{\Lambda },  \label{def}
\end{equation}%
where $(p^{\Lambda },q_{\Lambda })$ are $2n_{\upsilon }+2$ conserved charges
and $(L^{\Lambda },M_{\Lambda })$ are covariantly holomorphic sections
depending on moduli, where $L^{\Lambda }=e^{K/2}X^{\Lambda }$ and $%
M_{\Lambda }=e^{K/2}F_{\Lambda }$ \cite{Kallosh96}.

The equations for the double-extreme black holes in $N=2$ theory were
discussed \cite{Ferrara-1, Ferrara-2, Kallosh96, Levay06} and referred to as
\textquotedblleft stabilization equations\textquotedblright\ \cite{Behrndt,
Behrndt-2, Kallosh96, Levay06}. In \cite{Ferrara-1, Ferrara-2, Kallosh96,
Levay06},\ for supersymmetric black holes in $N=2$ theory near the horizon
the stabilization equations for $n_{\upsilon }$ complex moduli are given as
follows via the central charge $Z$
\begin{equation}
\left(
\begin{array}{c}
P^{\Lambda } \\
q_{\Lambda }%
\end{array}%
\right) =\left(
\begin{array}{c}
\mathbf{i}(Z^{\prime }L^{\Lambda }-ZL^{\prime \Lambda }) \\
\mathbf{i}(Z^{\prime }M_{\Lambda }-ZM_{\Lambda }^{\prime })%
\end{array}%
\right) ,  \label{motion-1}
\end{equation}%
where $\mathbf{i}$ is the imaginary unit and $Z^{\prime }$ is the complex
conjugate of the central charge $Z$. The stabilization equations are
covariant under symplectic transformations.

In \cite{Behrndt},\ to solve the stabilization equations\ on the
prepotential $F=\frac{X^{1}X^{2}X^{3}}{X^{0}}$, they first obtained the
matrix equation from the stabilization equations\ by eliminating $Z^{\prime
} $. Then, by solving the matrix equation the black hole entropy was first
derived in \cite{Behrndt, Behrndt-2, Kallosh-2021}. The black hole entropy
is a quartic homogeneous polynomial in the charges and invariant under $%
[SL(2,Z)]^{3}$ and under the STU triality.\

In \cite{Duff}, Duff related quantum information theory to the physics of
stringy black holes via the entropy of the extremal BPS STU black hole and
3-tangle for three qubits. Since then, the correspondence between pure
states of qubits in quantum information theory and black holes in string
theory has been established \cite{Duff, Duff08, Linde, Kallosh-2021,
Duff-review, Borsten, Meza}. Invoking the correspondence, Borsten et al.
\cite{Borsten} derived the SLOCC entanglement classification of four qubits,
and SLOCC and LU classifications of STU black holes were studied \cite%
{Linde, Li-1, Li-2}, where SLOCC\ (LU) equivalence for pure states of $n$
qubits is defined as follows. Let $|\psi \rangle $ and $|\psi ^{\prime
}\rangle $ be any pure states of $n$ qubits. Then, from \cite{Dur, Kraus-prl}%
, $|\psi ^{\prime }\rangle $ is SLOCC (LU) equivalent to $|\psi \rangle $ if
and only if there are local invertible (local unitary) operators $A_{1}$, $%
A_{2}$, $\cdots $, and $A_{n}$ ($U_{1}$, $U_{2}$, $\cdots $, and $U_{n}$)
such that $|\psi ^{\prime }\rangle =A_{1}\otimes A_{2}\otimes \cdots \otimes
A_{n}|\psi \rangle $ ($|\psi ^{\prime }\rangle =U_{1}\otimes U_{2}\otimes
\cdots \otimes U_{n}|\psi \rangle $).

In this paper, by directly solving the stabilization equations for the
solution of frozen moduli, we find linear relations between the complex
moduli. Via the linear relation, we derive the unique solution of frozen
moduli and reduce the stabilization equations.

\section{The stabilization equations}

Notation: In \cite{Behrndt}, they used $p^{i}$ for the magnetic charges and $%
z^{i}$ for the moduli, where $p^{i}$ and $z^{i}$\ are not\ exponential
functions. In this paper, we use $p_{i}$ for the magnetic charges and $z_{i}$
for the moduli.

In [4], they studied double-extreme black holes associated with the geometry
of the Calabi-Yau moduli with the prepotential and focused mostly on
STU-symmetric model $F=\frac{X^{1}X^{2}X^{3}}{X^{0}}$. A general static
spherically symmetric black hole solution depends on four electric charges
denoted $q_{0}$, $q_{1}$, $q_{2}$, $q_{3}$ and four magnetic charges\
denoted $p_{0}$, $p_{1}$, $p_{2}$, $p_{3}$ \cite{Duff}.

Let

\begin{equation}
P^{\Lambda }=\left(
\begin{array}{c}
p_{0} \\
p_{1} \\
p_{2} \\
p_{3}%
\end{array}%
\right) ;q_{\Lambda }=\left(
\begin{array}{c}
q_{0} \\
q_{1} \\
q_{2} \\
q_{3}%
\end{array}%
\right) ;X^{\Lambda }=\left(
\begin{array}{c}
1 \\
z_{1} \\
z_{2} \\
z_{3}%
\end{array}%
\right) ;F_{\Lambda }=\left(
\begin{array}{c}
-z_{1}z_{2}z_{3} \\
z_{2}z_{3} \\
z_{1}z_{3} \\
z_{1}z_{2}%
\end{array}%
\right) .
\end{equation}

Let $z_{i}^{\prime }$ be the complex conjugate of $z_{i}$. Let $\mu =\mathbf{%
i}e^{K/2}$ and $\mu ^{\prime }$ be the complex conjugate of $\mu $. Then, $%
\mu ^{\prime }=-\mu $ and
\begin{equation}
\mu Z^{\prime }=-(\mu Z)^{\prime }.
\end{equation}

In terms of special coordinates, from Eqs. (\ref{motion-1}, \ref{def}) and
via $L^{\Lambda }=e^{K/2}X^{\Lambda }$ and $M_{\Lambda }=e^{K/2}F_{\Lambda }$
the stabilization equations in Eq. (\ref{motion-1}) can be rewritten as \cite%
{Behrndt, Behrndt-2, Kallosh96}%
\begin{eqnarray}
P^{\Lambda } &=&\mu (Z^{\prime }X^{\Lambda }-ZX^{\prime ^{\Lambda }})
\label{stab-1} \\
q_{\Lambda } &=&\mu (Z^{\prime }F_{\Lambda }-ZF_{\Lambda }^{\prime })
\label{stab-2}
\end{eqnarray}

By eliminating $Z^{\prime }$ from Eqs. (\ref{stab-1}, \ref{stab-2}), they
obtained the following matrix equation \cite{Behrndt, Kallosh96}
\begin{equation}
X^{\Lambda }q_{\Sigma }-P^{\Lambda }F_{\Sigma }=\mathbf{i}e^{K/2}Z(X^{\prime
\Lambda }F_{\Sigma }-X^{\Lambda }F_{\Sigma }).  \label{matrix-1}
\end{equation}%
Eq. (\ref{matrix-1}) is a $4\times 4$ matrix equation in \cite{Behrndt}
while a $2\times 2$ matrix equation for the so-called $N=2$ axion-dilaton
model and a $4\times 4$ matrix equation of the STU model are discussed in
Sec. III and Sec. IV of \cite{Kallosh96}, respectively. Thus, from 16 (4)
entries of the matrix in Eq. (\ref{matrix-1})\ obtain 16 (4) equations,
respectively. By solving the matrix equation in Eq. (\ref{matrix-1}), they
derived the moduli, the black hole entropy, and the mass of the
double-extreme\ black hole \cite{Behrndt}.

In this paper, we directly solve the stabilization equations in Eqs. (\ref%
{stab-1}, \ref{stab-2}). From the stabilization equations in Eq. (\ref%
{stab-1}), obtain the following four equations.

\begin{eqnarray}
p_{0} &=&\mu (Z^{\prime }-Z)  \label{m-1} \\
p_{1} &=&\mu (Z^{\prime }z_{1}-Zz_{1}^{\prime })  \label{m-2} \\
p_{2} &=&\mu (Z^{\prime }z_{2}-Zz_{2}^{\prime })  \label{m-3} \\
p_{3} &=&\mu (Z^{\prime }z_{3}-Zz_{3}^{\prime })  \label{m-4}
\end{eqnarray}

From the stabilization equations in\ Eq. (\ref{stab-2}), obtain the
following four equations

\begin{eqnarray}
q_{0} &=&\mu (Zz_{1}^{\prime }z_{2}^{\prime }z_{3}^{\prime }-Z^{\prime
}z_{1}z_{2}z_{3})  \label{m-5} \\
q_{1} &=&\mu (Z^{\prime }z_{2}z_{3}-Zz_{2}^{\prime }z_{3}^{\prime })\text{ }
\label{m-6} \\
q_{2} &=&\mu (Z^{\prime }z_{1}z_{3}-Zz_{1}^{\prime }z_{3}^{\prime })
\label{m-7} \\
q_{3} &=&\mu (Z^{\prime }z_{1}z_{2}-Zz_{1}^{\prime }z_{2}^{\prime })
\label{m-8}
\end{eqnarray}

We next solve the solutions of the complex moduli $z_{1},z_{2},$ and $z_{3}$
from the Eqs. (\ref{m-1}-\ref{m-8}). It is trivial to see the complex $\mu $
and the complex central charge $Z$ occur in Eqs. (\ref{m-1}-\ref{m-8}).

\subsection{Calculating $\protect\mu Z$}

Calculating $z_{1}\ast $ Eq. (\ref{m-1})-Eq. (\ref{m-2}), obtain
\begin{equation}
\text{ }p_{0}z_{1}^{\prime }-p_{1}=\mu Z^{\prime }\left( z_{1}^{\prime
}-z_{1}\right) .  \label{y1}
\end{equation}%
Then,
\begin{equation}
\mu Z=\frac{p_{0}z_{1}-p_{1}}{\left( z_{1}^{\prime }-z_{1}\right) }.
\label{y2}
\end{equation}

$\allowbreak $Calculating $z_{2}^{\prime }\ast $ Eq. (\ref{m-1})-Eq. (\ref%
{m-3}), obtain

\begin{equation}
\mu Z=\frac{p_{0}z_{2}-p_{2}}{\left( z_{2}^{\prime }-z_{2}\right) }.
\label{y4}
\end{equation}

Calculating $z_{3}^{\prime }\ast $ Eq. (\ref{m-1})-Eq. (\ref{m-4}), obtain

\begin{equation}
\mu Z=\frac{p_{0}z_{3}-p_{3}}{\left( z_{3}^{\prime }-z_{3}\right) }.
\label{y6}
\end{equation}

\section{The linear relations derived from stabilization equations}

Let $\omega _{1}=(p_{2}p_{3}-p_{0}q_{1})$, $\omega
_{2}=(p_{1}p_{3}-p_{0}q_{2})$, and $\omega _{3}=(p_{1}p_{2}-p_{0}q_{3})$. It
is clear that $\omega _{i}$, $i=1,2,3$, are real.

\subsection{The linear relation of $z_{1}$ and $z_{3}$}

Calculating $z_{1}\ast $ Eq. (\ref{m-3})-Eq. (\ref{m-8}), obtain

\begin{eqnarray}
p_{2}z_{1}-q_{3} &=&z_{2}^{\prime }(p_{0}z_{1}-p_{1}),  \label{inv-3} \\
z_{2}^{\prime } &=&\frac{p_{2}z_{1}-q_{3}}{p_{0}z_{1}-p_{1}}.  \label{H1}
\end{eqnarray}

$\allowbreak $Calculating $z_{3}\ast $ Eq. (\ref{m-3})-Eq. (\ref{m-6}), via
Eq. (\ref{y6}) obtain

\begin{equation}
z_{2}^{\prime }=\frac{p_{2}z_{3}-q_{1}}{\left( -p_{3}+p_{0}z_{3}\right) }.
\label{H2}
\end{equation}

From Eqs. (\ref{H1}, \ref{H2}), obtain the following linear relation:
\begin{equation}
-p_{1}q_{1}+p_{3}q_{3}-\omega _{1}z_{1}+\omega _{3}z_{3}=0.  \label{H3}
\end{equation}

One can see that in Eq. (\ref{H3}) the coefficients $\omega _{1}$, $\omega
_{3}$ and $-p_{1}q_{1}+p_{3}q_{3}$ depend on the charges and are real.

\subsection{The linear relation of $z_{1}$ and $z_{2}$}

$\allowbreak $Calculating $z_{1}\ast $ Eq. (\ref{m-4})-Eq. (\ref{m-7}), via
Eq. (\ref{y2}) obtain

\begin{eqnarray}
p_{3}z_{1}-q_{2} &=&-z_{3}^{\prime }\left( p_{1}-p_{0}z_{1}\right) , \\
z_{3}^{\prime } &=&\frac{p_{3}z_{1}-q_{2}}{p_{0}z_{1}-p_{1}}.  \label{D1}
\end{eqnarray}

Calculating $z_{2}\ast $ Eq. (\ref{m-4})-Eq. (\ref{m-6}), via Eq. (\ref{y4})
obtain%
\begin{equation}
\text{ }z_{3}^{\prime }=\frac{p_{3}z_{2}-q_{1}}{\left(
-p_{2}+p_{0}z_{2}\right) }.  \label{D2}
\end{equation}

From Eqs. (\ref{D1}, \ref{D2}), obtain the following linear relation:
\begin{equation}
-p_{1}q_{1}+p_{2}q_{2}-\omega _{1}z_{1}+\omega _{2}z_{2}=0.  \label{D3-1}
\end{equation}

One can see that in Eq. (\ref{D3-1}) the coefficients $\omega _{1}$, $\omega
_{2}$ and $-p_{1}q_{1}+p_{2}q_{2}$ depend on the charges and are real.

\subsection{The linear relation of $z_{2}$ and $z_{3}$}

$\allowbreak $Calculating $z_{3}\ast $ Eq. (\ref{m-2})-Eq. (\ref{m-7}), via
Eq. (\ref{y6}) obtain

\begin{eqnarray}
p_{1}z_{3}-q_{2} &=&z_{1}^{\prime }\left( -p_{3}+p_{0}z_{3}\right) , \\
z_{1}^{\prime } &=&\frac{p_{1}z_{3}-q_{2}}{\left( -p_{3}+p_{0}z_{3}\right) }.
\label{G1}
\end{eqnarray}

$\allowbreak $Calculating $z_{2}\ast $ Eq. (\ref{m-2})-Eq. (\ref{m-8}), via
Eq. (\ref{y4}) obtain

\begin{equation}
z_{1}^{\prime }=\frac{p_{1}z_{2}-q_{3}}{\left( -p_{2}+p_{0}z_{2}\right) }.
\label{G2}
\end{equation}

From Eqs. (\ref{G1}, \ref{G2}), obtain the following linear relation:

\begin{equation}
p_{2}q_{2}-p_{3}q_{3}+\omega _{2}z_{2}-\omega _{3}z_{3}=0.  \label{G3}
\end{equation}

One can see that in Eq. (\ref{G3}) the coefficients $\omega _{2}$, $\omega
_{3}$ and $p_{2}q_{2}-p_{3}q_{3}$ depend on the charges and are real. The
linear relation in Eq. (\ref{G3}) can also be obtained from the linear
relations in Eqs. (\ref{D3-1}), \ref{H3}).

It is known that so far, the linear relations between any two of the three
complex moduli of the STU model have not made their appearance in the
previous literatures.

\subsection{The property of the moduli}

From Eqs. (\ref{H3}, \ref{D3-1}), it is easy to know

\begin{equation}
\left( z_{1}-z_{1}^{\prime }\right) \omega _{1}=(z_{2}-z_{2}^{\prime
})\omega _{2}=(z_{3}-z_{3}^{\prime })\omega _{3},  \label{ima}
\end{equation}

Eq. (\ref{ima}) can be rewritten as the follow:

\begin{equation}
im(z_{1}):im(z_{2}):im(z_{3})=\omega _{2}\omega _{3}:\omega _{1}\omega
_{3}:\omega _{1}\omega _{2},  \label{ima-2}
\end{equation}%
where $im(z_{1})$ is the imaginary part of the complex number $z_{1}$.

One can see that there are the linear relations between any two of the three
complex moduli and the linear relations imply the imaginary parts of the
moduli must satisfy Eqs. (\ref{ima}, \ref{ima-2}). Eq. (\ref{ima}) claims $%
\left( z_{i}-z_{i}^{\prime }\right) \omega _{i}$, $i=1,2,3$, are invariant
under the transformations $z_{1}\Longleftrightarrow z_{2}\Longleftrightarrow
z_{3}$, $p_{1}\Longleftrightarrow p_{2}\Longleftrightarrow p_{3}$ and $%
q_{1}\Longleftrightarrow q_{2}\Longleftrightarrow q_{3}$. That is, $\left(
z_{i}-z_{i}^{\prime }\right) \omega _{i}$, $i=1,2,3$, are invariant under
the transformations $1\Longleftrightarrow 2\Longleftrightarrow 3$ for the
indices, i.e. the STU\ triality.

\section{The linear relations determine the uniqueness of solutions of
moduli $z_{i}$}

\subsection{Calculating $z_{2}z_{3}$}

Calculating $z_{1}^{\prime }\ast $ Eq. (\ref{m-6})+Eq. (\ref{m-5}), from Eq.
(\ref{y2}) obtain
\begin{equation}
z_{2}z_{3}=\frac{q_{1}z_{1}^{\prime }+q_{0}}{p_{0}z_{1}^{\prime }-p_{1}}.
\label{W1}
\end{equation}

\subsection{The equation which $z_{1}$ satisfies}

Let
\begin{eqnarray}
B_{1} &=&\left( p_{0}q_{0}-p_{1}q_{1}+p_{2}q_{2}+p_{3}q_{3}\right)
,C_{1}=(p_{1}q_{0}+q_{2}q_{3}) \\
B_{2}
&=&(p_{0}q_{0}+p_{1}q_{1}-p_{2}q_{2}+p_{3}q_{3}),C_{2}=(p_{2}q_{0}+q_{1}q_{3})
\\
B_{3}
&=&(p_{0}q_{0}+p_{1}q_{1}+p_{2}q_{2}-p_{3}q_{3}),C_{3}=(p_{3}q_{0}+q_{1}q_{2})
\end{eqnarray}

From Eqs. (\ref{H1}, \ref{D1}, \ref{W1}), obtain
\begin{equation}
\omega _{1}z_{1}^{2}-B_{1}z_{1}+C_{1}=0.  \label{s1}
\end{equation}

\subsection{The linear relations determine the uniqueness of solutions of
moduli $z_{i}$}

\subsubsection{The imaginary parts of the moduli $z_{i}$ have the same sign}

From Eq. (\ref{s1}), we can obtain the value of $z_{1}$ as follows. \
\begin{equation}
z_{1}=\frac{B_{1}}{2\omega _{1}}+\mathbf{i}\delta \frac{\sqrt{W}}{2\omega
_{1}},\delta =\pm 1,  \label{md-1}
\end{equation}%
where
\begin{eqnarray}
W &=&-(p_{0}q_{0}+p_{1}q_{1}+p_{2}q_{2}+p_{3}q_{3})^{2}  \nonumber \\
&&+4(p_{1}q_{1}p_{2}q_{2}+p_{1}q_{1}p_{3}q_{3}+p_{3}q_{3}p_{2}q_{2})-4p_{0}q_{1}q_{2}q_{3}+4q_{0}p_{1}p_{2}p_{3}.
\end{eqnarray}

Note that $W>0$ and $W$ is referred to as the black hole entropy \cite%
{Behrndt}.\ From Eq. (\ref{md-1}) and the linear relations in Eqs. (\ref%
{D3-1}, \ref{H3}), obtain
\begin{eqnarray}
z_{2} &=&\frac{B_{2}}{2\omega _{2}}+\mathbf{i}\delta \frac{%
\sqrt{W}}{2\omega _{2}},  \label{md-2} \\
z_{3} &=&\frac{B_{3}}{2\omega _{3}}+\mathbf{i}\delta \frac{\sqrt{W}}{2\omega
_{3}}.  \label{md-3}
\end{eqnarray}

From Eqs. (\ref{md-1}, \ref{md-2}, \ref{md-3}), one can see that the linear
relations imply that the imaginary parts of the moduli $z_{i}$ have the same
sign $\delta $.

\subsubsection{The uniqueness of the solutions of the moduli}

\textit{Theorem 1}. The unique solutions of the moduli $z_{j}$ are

\begin{equation}
z_{j}=\frac{B_{j}}{2\omega _{j}}-\mathbf{i}\frac{\sqrt{W}}{2\omega _{j}}%
,j=1,2,3.  \label{md}
\end{equation}

Proof. From the corresponding K\"{a}hler potential $K$ in Eq. (19) in \cite%
{Behrndt}, one knows that $K$ is real and

\begin{equation}
e^{-K}=-\mathbf{i}(z_{1}-z_{1}^{\prime })(z_{2}-z_{2}^{\prime
})(z_{3}-z_{3}^{\prime }).
\end{equation}

Thus, via Eqs. (\ref{md-1}, \ref{ima}) or Eqs. (\ref{md-1}, \ref{md-2}, \ref%
{md-3}), obtain
\begin{equation}
e^{-K}=-\frac{\delta ^{3}\left( \sqrt{W}\right) ^{3}}{\omega _{1}\omega
_{2}\omega _{3}}.  \label{ten-1}
\end{equation}

It is known that $\omega _{1}\allowbreak \omega _{2}\omega _{3}>0$ \cite%
{Behrndt}. Clearly, $e^{-K}>0$ if and only if $\delta =-1$. Of course, it is
natural to require that $e^{-K}>0$ because $K$ is real. Then, when $\delta
=-1$,
\begin{equation}
e^{-K}=\frac{\left( \sqrt{W}\right) ^{3}}{\omega _{1}\omega _{2}\omega _{3}}.
\label{ten-2}
\end{equation}

From Eqs. (\ref{md-1}, \ref{md-2}, \ref{md-3}) and $\delta =-1$, clearly Eq.
(\ref{md}) holds.

Q.E.D.

\textit{Remark 1.}

Eq. (37) of \cite{Behrndt} gave the all possible solutions of moduli $z_{j}$
as follows.
\begin{equation}
z_{j}^{\pm }=\frac{B_{j}}{2\omega _{j}}\pm \mathbf{i}\frac{\sqrt{W}}{2\omega
_{j}},j=1,2,3.  \label{mod-0}
\end{equation}%
One can see there are eight possible cases $\{z_{1}^{\pm },z_{2}^{\pm
},z_{3}^{\pm }\}$. We can demonstrate that the K\"{a}hler potential\ $%
e^{-K}>0$ \ for the following four cases: Case 1.$\{z_{1}^{-}$, $z_{2}^{-}$,
$z_{3}^{-}\}$, Case 2. $\{z_{1}^{-}$, $z_{2}^{+}$, $z_{3}^{+}\}$, Case 3. $%
\{z_{1}^{+}$, $z_{2}^{-}$, $z_{3}^{+}\}$, and Case 4. $\{z_{1}^{+}$, $%
z_{2}^{+}$, $z_{3}^{-}\}$, \ while $e^{-K}<0$ for Case 5. $\{z_{1}^{-}$, $%
z_{2}^{-}$, $z_{3}^{+}\}$, Case 6. $\{z_{1}^{-}$, $z_{2}^{+}$, $z_{3}^{-}\}$%
, Case 7. $\{z_{1}^{+}$, $z_{2}^{-}$, $z_{3}^{-}\}$, and Case 8. $%
\{z_{1}^{+} $, $z_{2}^{+}$, $z_{3}^{+}\}$.

For example, we can show that $e^{-K}>0$ for Case 2 as follows. For Case 2, $%
z_{1}^{-}=\frac{B_{1}}{2\omega _{1}}-\mathbf{i}\frac{\sqrt{W}}{2\omega _{1}}$
(for which the sign for the imaginary part is negative) and $z_{j}^{+}=\frac{%
B_{j}}{2\omega _{j}}+\mathbf{i}\frac{\sqrt{W}}{2\omega _{j}},$ $j=2,3$, (for
which the signs for the imaginary parts are positive). Thus, $e^{-K}=-%
\mathbf{i}(z_{1}^{-}-z_{1}^{-\prime })(z_{2}^{+}-z_{2}^{+\prime
})(z_{3}^{+}-z_{3}^{+\prime })=\frac{\left( \sqrt{W}\right) ^{3}}{\omega
_{1}\allowbreak \omega _{2}\omega _{3}}>0$.

Similarly, we can show that $e^{-K}>0$ for Cases 1, 3, and 4. Note that for
Case 1, the signs for the imaginary parts of the moduli $z_{j}$, $j=1,2,3$,
are negative while for any one of Cases 2-4, the signs for imaginary parts
of just two moduli are positive.

Of course, it is natural to require $e^{-K}>0$. Therefore, Cases 5-8 are not
solutions of the stabilization equations. As mentioned above, for Cases 1-4,
$e^{-K}>0$. Then, which of Cases 1-4 are true solutions for the
stabilization equations? Is there a unique solution for the stabilization
equations? The authors of \cite{Behrndt}\ said that for the K\"{a}hler
potential $e^{-K}$ to be positive, they had to pick up only one choice of
sign for each imaginary part of the moduli $z_{j}$, $j=1,2,3$, in Eq. (\ref%
{mod-0}), it has to be negative. That is, for the K\"{a}hler potential $%
e^{-K}$ to be positive, they had to pick up Case 1. From the above
discussion, it is known that for $e^{-K}$ to be positive, clearly it is not
necessary to require the sign for each imaginary part of moduli $z_{j}$, $%
j=1,2,3$, to be negative because also $e^{-K}>0$ for Cases 2-4. Theorem 1 in
this paper deduces that Case 1 is a unique choice via the linear relations.

\textit{Remark 2.}

It is trivial to know that the linear relations in Eqs. (\ref{H3}, \ref{D3-1}%
, \ref{G3}) can also be derived from Case 8: $z_{j}^{+}$, $j=1,2,3$, for
which $e^{-K}<0$ and Case 1: $z_{j}^{-}$, $j=1,2,3$, for which $e^{-K}>0$,
respectively. Note that $z_{j}^{+}$, $j=1,2,3$, are not solutions of the
stabilization equations.

From $z_{1}$ in Eq. (\ref{md-1}), via the linear relations in Eqs. (\ref%
{D3-1}, \ref{H3}) we derive $z_{2}$ and $z_{3}$ in Eqs. (\ref{md-2}, \ref%
{md-3}). Ref. Section 4.3.1. Conversely, Eq. (\ref{md}) implies the linear
relations. Thus, it is clear that the following Proposition 1 holds.

\textit{Proposition 1}.

The linear relations in Eqs. (\ref{H3}, \ref{D3-1}) and the solution of $%
z_{1}$ in Eq. (\ref{md}) together are equivalent to the solutions of the
moduli $z_{j}$, $j=1,2,3$, in Eq. (\ref{md}).

\subsection{Calculating the central charge $Z$ and $\protect\mu Z$\ via the
linear relations}

Note that the central charge was not derived in \cite{Behrndt}, though $%
ZZ^{\prime }$ was calculated therein. By the definition of the central
charge in Eq. (\ref{def}) and via Eqs. (\ref{md}, \ref{ten-2}) , obtain the
central charge

\begin{equation}
Z=\frac{\sqrt[4]{W}}{2\sqrt{\omega _{1}\omega _{2}\omega _{3}}}\left(
-p_{0}\sum_{i=0}^{3}p_{i}q_{i}+2p_{1}p_{2}p_{3}+\mathbf{i}\sqrt{W}%
p_{0}\right) .  \label{charge-1}
\end{equation}

Via Eqs. (\ref{y2}, \ref{ten-2}, \ref{md-1}) and $\mu =\mathbf{i}e^{K/2}$,
we can also derive the central charge. From (\ref{charge-1}),\ It is easy to
verify $|Z|^{2}=\sqrt{W}$ and thus, we verify the formula for the mass/area
for the black hole \cite{Behrndt}.

Via the linear relations, one can see that the central charge is the
function of the moduli $z_{1}$ and it can be written as the following.

\begin{equation}
Z=-\mathbf{i}e^{-K/2}\frac{p_{0}z_{1}-p_{1}}{\left( z_{1}^{\prime
}-z_{1}\right) }=-\frac{W^{1/4}\omega _{1}}{\sqrt{\omega _{1}\omega
_{2}\omega _{3}}}(p_{0}z_{1}-p_{1}).
\end{equation}

Note that $\mu Z$ appears in each one of the eight stabilization equations.
From Eqs. (\ref{ten-2},\ref{charge-1}), a calculation yields
\begin{eqnarray}
e^{K/2} &=&\frac{\sqrt{\omega _{1}\omega _{2}\omega _{3}}}{W^{3/4}},\mu =%
\mathbf{i}e^{K/2}, \\
\mu Z &=&\frac{1}{2}\left( \mathbf{-}p_{0}+\mathbf{i}\frac{1}{\sqrt{W}}%
(-p_{0}\sum_{i=0}^{3}p_{i}q_{i}+2p_{1}p_{2}p_{3})\right) .  \label{inv-1}
\end{eqnarray}

From Eq. (\ref{inv-1}), one can see that $\mu Z$ is invariant under the
transformations $p_{1}\Longleftrightarrow p_{2}\Longleftrightarrow p_{3}$
and $q_{1}\Longleftrightarrow q_{2}\Longleftrightarrow q_{3}$ and the real
part of $\mu Z$ is $\mathbf{-}p_{0}/2$.

\section{The stabilization equations can be replaced equivalently with the
simple and intuitive conditions}

The stabilization equations were studied in previous papers \cite{Ferrara-1,
Ferrara-2, Kallosh96, Behrndt, Behrndt-2, Kallosh-2021, Levay06}. It is
valuable to reduce the stabilization equations because it has had a
significant impact. Note that a complex constraint among the symplectic
sections can also be used to reduce the stabilization equations \cite%
{Ceresole-2}.

\textit{Theorem 2}. The eight stabilization equations are equivalent to the
following four equations

\begin{eqnarray}
\omega _{2}z_{2} &=&p_{1}q_{1}-p_{2}q_{2}+\omega _{1}z_{1},  \label{cd-1} \\
\omega _{3}z_{3} &=&p_{1}q_{1}-p_{3}q_{3}+\omega _{1}z_{1},  \label{cd-2} \\
\mu Z &=&\frac{p_{0}z_{1}-p_{1}}{\left( z_{1}^{\prime }-z_{1}\right) },
\label{cd-3} \\
z_{2}z_{3} &=&\frac{q_{1}z_{1}^{\prime }+q_{0}}{p_{0}z_{1}^{\prime }-p_{1}},
\label{ccd-4}
\end{eqnarray}%
where Eqs. ( \ref{cd-1}, \ref{cd-2}) are the linear relations in Eqs. (\ref%
{D3-1}, \ref{H3}), Eq. (\ref{cd-3}) is Eq. (31) in \cite{Behrndt}, and Eq. (%
\ref{ccd-4}) is Eq. (30) in \cite{Behrndt}.

\textit{Proof}. We have demonstrated how to derive Eqs. (\ref{cd-1}-\ref%
{ccd-4}) from the stabilization equations. We next derive the stabilization
equations from Eqs. (\ref{cd-1}-\ref{ccd-4}) in Appendix A.

\textit{Theorem 3}. The eight stabilization equations are equivalent to the
following three equations

\begin{eqnarray}
\omega _{2}z_{2} &=&p_{1}q_{1}-p_{2}q_{2}+\omega _{1}z_{1},  \label{rel-2} \\
\omega _{3}z_{3} &=&p_{1}q_{1}-p_{3}q_{3}+\omega _{1}z_{1},  \label{rel-3} \\
z_{1} &=&\frac{B_{1}}{2\omega _{1}}-\mathbf{i}\frac{\sqrt{W}}{2\omega _{1}},
\label{z1} \\
&&  \nonumber
\end{eqnarray}%
where Eqs. ( \ref{rel-2}, \ref{rel-3}) are the linear relations in Eqs. (\ref%
{D3-1}, \ref{H3}) and Eq. (\ref{z1}) is the value of the moduli $z_{1}$.

\textit{Proof.} We have demonstrated how to derive Eqs. ( \ref{rel-2}, \ref%
{rel-3}, \ref{z1}) from the stabilization equations. We next derive the
stabilization equations from Eqs. ( \ref{rel-2}, \ref{rel-3}, \ref{z1}) in
Appendix B.

From Proposition 1 and Theorem 3 we have the following.

\textit{Theorem 4}. The eight stabilization equations hold if and only if
the following
\begin{equation}
z_{j}=\frac{B_{j}}{2\omega _{j}}-\mathbf{i}\frac{\sqrt{W}}{2\omega _{j}}%
,j=1,2,3.  \label{mod-1}
\end{equation}

That is, the eight stabilization equations are equivalent to the solutions
of $z_{j}$, $j=1,2,3$.

Theorem 4 means that $z_{j}$, $j=1,2,3$, in Eq. (\ref{mod-1}) satisfy the
eight stabilization equations. Thus, it verifies that $z_{j}$, $j=1,2,3$, in
Eq. (\ref{mod-1}) are true solutions of the stabilization equations.

\textit{Remark 3}. Note that each of the eight equations in Eqs. (\ref{m-1}-%
\ref{m-8}) is given via $\mu Z$, where $\mu =\mathbf{i}e^{K/2}$ and the
central charge $Z$ depends on the complex moduli and the charges. $%
z_{1}z_{2}z_{3}$\ appears in Eq. (\ref{m-5}) and $z_{i}z_{j}$ appear in Eqs.
(\ref{m-6}-\ref{m-8}).

Comparatively, $\mu Z$ does not appear in Eqs. ( \ref{rel-2}, \ref{rel-3}, %
\ref{ccd-4}, \ref{mod-1}). One can see that the coefficients of the moduli $%
z_{1},z_{2},$ and $z_{3}$ in the linear relations in Eqs. (\ref{rel-2}, \ref%
{rel-3}) depend on the charges and are real. Clearly, Eqs. (\ref{cd-1}-\ref%
{ccd-4}), Eqs. (\ref{rel-2}-\ref{z1}), and Eq. (\ref{mod-1}) are much
simpler and more intuitive than Eqs. (\ref{m-1}-\ref{m-8}).

\section{Reduce stabilization equations via the STU triality}

We show that $\mu Z$ is invariant under the STU triality in Section 4.4.
Under the transformations $z_{1}\Longleftrightarrow z_{2}$, $%
p_{1}\Longleftrightarrow p_{2}$, and $q_{1}\Longleftrightarrow q_{2}$, i.e.
transformations $1\Longleftrightarrow 2$ for the indices, then Eq. (\ref{m-2}%
) $\Longleftrightarrow $Eq. (\ref{m-3}) and Eq. (\ref{m-6}) $%
\Longleftrightarrow $Eq. (\ref{m-7}), respectively. Under the transformation
$z_{1}\Longleftrightarrow z_{3}$, $p_{1}\Longleftrightarrow p_{3}$, and $%
q_{1}\Longleftrightarrow q_{3}$, i.e. transformations $1\Longleftrightarrow
3 $ for the indices, then Eq. (\ref{m-2}) $\Longleftrightarrow $Eq. (\ref%
{m-4}) and Eq. (\ref{m-6}) $\Longleftrightarrow $Eq. (\ref{m-8}),
respectively. Thus, via the STU triality the stabilization equations in Eqs.
(\ref{m-1}-\ref{m-8}) can be reduced. For example, the eight stabilization
equations in Eqs. (\ref{m-1}-\ref{m-8}) can be replaced with the four
equations in Eqs. (\ref{m-1}, \ref{m-2}, \ref{m-5}, \ref{m-6}) under the STU
triality.

By applying the STU triality to the linear relation in Eq. (\ref{H3}), we
can obtain the linear relations in Eqs. (\ref{D3-1}, \ref{G3}). Similarly,
by applying the STU triality to $z_{1}$ in Eq. (\ref{mod-1}), obtain $z_{2}$
and $z_{3}$ in Eq. (\ref{mod-1}). Thus, via the STU triality, \ we can
rephrase Theorems 2-4 in Table 1.

Table 1. Rephrase Theorems 2-4 via the STU triality

\begin{tabular}{|l|l|}
\hline
& The stabilization equations are equivalent to \\ \hline
Theorem 2 & $\omega _{2}z_{2}=p_{1}q_{1}-p_{2}q_{2}+\omega _{1}z_{1},\mu Z=%
\frac{p_{0}z_{1}-p_{1}}{\left( z_{1}^{\prime }-z_{1}\right) },z_{2}z_{3}=%
\frac{q_{1}z_{1}^{\prime }+q_{0}}{p_{0}z_{1}^{\prime }-p_{1}}$ \\ \hline
Theorem 3 & $\omega _{2}z_{2}=p_{1}q_{1}-p_{2}q_{2}+\omega _{1}z_{1},z_{1}=%
\frac{B_{1}}{2\omega _{1}}-\mathbf{i}\frac{\sqrt{W}}{2\omega _{1}}$ \\ \hline
Theorem 4 & $z_{1}=\frac{B_{1}}{2\omega _{1}}-\mathbf{i}\frac{\sqrt{W}}{%
2\omega _{1}}$ \\ \hline
\end{tabular}

\section{Summary}

The black hole entropy was derived by solving the matrix equation obtained
from the stabilization equations for the solution of frozen moduli \cite%
{Behrndt}. In this paper, by directly solving the stabilization equations
for the solution of frozen moduli, we find linear relations between any two
of the three complex moduli, where the coefficients of three moduli depend
on the charges and are real. The linear relations were not discussed in
previous papers.

Via the linear relations we derive the unique solutions of frozen moduli and
reduce the stabilization equations. We demonstrate that the stabilization
equations in Eqs. (\ref{m-1}-\ref{m-8})\ can be replaced equivalently with
three equations in Eqs. ( \ref{rel-2}-\ref{z1}), four equations in Eqs. (\ref%
{cd-1}-\ref{ccd-4}), and the solutions of three moduli $z_{1},z_{2}$ and $%
z_{3}$ in Eq. (\ref{mod-1}), respectively. Eqs. ( \ref{rel-2}-\ref{z1}),
Eqs. (\ref{cd-1}-\ref{ccd-4}), and Eq. (\ref{mod-1}) are much simpler and
more intuitive than the stabilization equations in Eqs. (\ref{m-1}-\ref{m-8}%
). These discussions may be useful to study the stabilization equations for
eight and 16 charges \cite{Meza}.

Data Availability Statement:

No data associated in the manuscript.

\section{Appendix A Proof of Theorem 2}

\renewcommand{\theequation}{A\arabic{equation}} \setcounter{equation}{0}

From the linear relation in Eq. (\ref{cd-1}) and Eq. (\ref{cd-3}), obtain

\begin{equation}
\mu Z=\frac{p_{0}z_{2}-p_{2}}{\left( z_{2}^{\prime }-z_{2}\right) }.
\label{cd5}
\end{equation}

From the linear relation in Eq. (\ref{cd-2}) and Eq. (\ref{cd-3}), obtain

\begin{equation}
\mu Z=\frac{p_{0}z_{3}-p_{3}}{\left( z_{3}^{\prime }-z_{3}\right) }.
\label{cd6}
\end{equation}

From Eqs. (\ref{cd-3}, \ref{cd5}, \ref{cd6}), obtain
\begin{eqnarray}
\mu Z^{\prime } &=&\frac{p_{0}z_{1}^{\prime }-p_{1}}{\left( z_{1}^{\prime
}-z_{1}\right) },  \label{cd7} \\
\mu Z^{\prime } &=&\frac{p_{0}z_{2}^{\prime }-p_{2}}{\left( z_{2}^{\prime
}-z_{2}\right) },  \label{cd8} \\
\mu Z^{\prime } &=&\frac{p_{0}z_{3}^{\prime }-p_{3}}{\left( z_{3}^{\prime
}-z_{3}\right) }.  \label{cd9}
\end{eqnarray}

Then, it is easy to derive Eqs. (\ref{m-1}, \ref{m-2}) via Eqs. (\ref{cd-3}, %
\ref{cd7}), Eq. (\ref{m-3}) via Eqs. (\ref{cd5}, \ref{cd8}), and Eq. (\ref%
{m-4}) via Eqs. (\ref{cd6}, \ref{cd9}), respectively.

From Eq. (\ref{ccd-4}), obtain
\begin{equation}
z_{2}^{\prime }z_{3}^{\prime }=\frac{q_{1}z_{1}+q_{0}}{p_{0}z_{1}-p_{1}}.
\label{cd10}
\end{equation}

Then, one can see that Eqs. (\ref{m-5}, \ref{m-6}) hold via Eqs. (\ref{cd-3}%
, \ref{cd7}, \ref{ccd-4}, \ref{cd10}).

Let us derive Eq. (\ref{m-7}). Via the linear relation in Eq. ( \ref{cd-1}),
obtain
\begin{eqnarray}
\omega _{1}z_{1}z_{3} &=&\omega _{2}z_{2}z_{3}-(p_{1}q_{1}-p_{2}q_{2})z_{3},
\label{cd11} \\
\omega _{1}z_{1}^{\prime }z_{3}^{\prime } &=&\omega _{2}z_{2}^{\prime
}z_{3}^{\prime }-(p_{1}q_{1}-p_{2}q_{2})z_{3}^{\prime }.  \label{cd12}
\end{eqnarray}

Then, via Eqs. (\ref{cd11}, \ref{cd12}, \ref{m-4}, \ref{m-6}), a calculation
yields
\begin{eqnarray}
&&\omega _{1}(\mu (Z^{\prime }z_{1}z_{3}-Zz_{1}^{\prime }z_{3}^{\prime }))
\nonumber \\
&=&\mu Z^{\prime }(\omega _{2}z_{2}z_{3}-(p_{1}q_{1}-p_{2}q_{2})z_{3})-\mu
Z(\omega _{2}z_{2}^{\prime }z_{3}^{\prime
}-(p_{1}q_{1}-p_{2}q_{2})z_{3}^{\prime })  \nonumber \\
&=&\omega _{2}(\mu Z^{\prime }z_{2}z_{3}-\mu Zz_{2}^{\prime }z_{3}^{\prime
})+(p_{1}q_{1}-p_{2}q_{2})(\mu Zz_{3}^{\prime }-\mu Z^{\prime }z_{3})
\nonumber \\
&=&\omega _{2}q_{1}-(p_{1}q_{1}-p_{2}q_{2})p_{3}  \nonumber \\
&=&\omega _{1}q_{2}.  \label{cd13}
\end{eqnarray}

Clearly, Eq. (\ref{m-7}) holds via Eq. (\ref{cd13}).

Let us derive Eq. (\ref{m-8}). Via the linear relation in Eq. ( \ref{cd-2}),
obtain
\begin{eqnarray}
\omega _{1}z_{1}z_{2} &=&\omega _{3}z_{2}z_{3}-(p_{1}q_{1}-p_{3}q_{3})z_{2},
\label{cd14} \\
\omega _{1}z_{1}^{\prime }z_{2}^{\prime } &=&\omega _{3}z_{2}^{\prime
}z_{3}^{\prime }-(p_{1}q_{1}-p_{3}q_{3})z_{2}^{\prime }.  \label{cd15}
\end{eqnarray}

Then, via Eqs. (\ref{cd14}, \ref{cd15}, \ref{m-3}, \ref{m-6}), a calculation
yields
\begin{eqnarray}
&&\omega _{1}(\mu (Z^{\prime }z_{1}z_{2}-Zz_{1}^{\prime }z_{2}^{\prime }))
\nonumber \\
&=&\mu Z^{\prime }(\omega _{3}z_{2}z_{3}-(p_{1}q_{1}-p_{3}q_{3})z_{2})-\mu
Z(\omega _{3}z_{2}^{\prime }z_{3}^{\prime
}-(p_{1}q_{1}-p_{3}q_{3})z_{2}^{\prime })  \nonumber \\
&=&\omega _{3}(\mu Z^{\prime }z_{2}z_{3}-\mu Zz_{2}^{\prime }z_{3}^{\prime
})+(p_{1}q_{1}-p_{3}q_{3})(\mu Zz_{2}^{\prime }-\mu Z^{\prime }z_{2})
\nonumber \\
&=&\omega _{3}q_{1}-(p_{1}q_{1}-p_{3}q_{3})p_{2}  \nonumber \\
&=&(p_{2}p_{3}-p_{0}q_{1})q_{3}.  \label{cd16}
\end{eqnarray}

Clearly, Eq. (\ref{m-8}) holds via Eq. (\ref{cd16}). Q.E.D.

\section{Appendix B. The proof of Theorem 3}

\renewcommand{\theequation}{B\arabic{equation}} \setcounter{equation}{0}

From Eqs. (\ref{rel-2}, \ref{rel-3}, \ref{z1}), obtain
\begin{equation}
z_{j}=\frac{B_{j}}{2\omega _{j}}-\mathbf{i}\frac{\sqrt{W}}{2\omega _{j}}%
,j=1,2,3.  \label{sol-zi}
\end{equation}

From Eq. (\ref{sol-zi}), obtain
\begin{equation}
(z_{j}-z_{j}^{\prime })=\frac{-\mathbf{i}\sqrt{W}}{\omega _{j}},j=1,2,3.
\label{imig-1}
\end{equation}

Via Eq. (\ref{imig-1}), a calculation yields
\begin{eqnarray}
e^{-K} &=&\frac{\left( \sqrt{W}\right) ^{3}}{\omega _{1}\allowbreak \omega
_{2}\omega _{3}},  \label{K} \\
\mu &=&\mathbf{i}e^{K/2}=\mathbf{i}\frac{(\omega _{1}\allowbreak \omega
_{2}\omega _{3})^{1/2}}{W^{3/4}}.  \label{mu-1}
\end{eqnarray}

Again, by the definition of the central change in Eq. (\ref{def}) and via
Eqs. (\ref{imig-1}, \ref{K}), a tedious and straightforward calculation
yields

\begin{equation}
Z=\frac{\sqrt[4]{W}}{2\sqrt{\omega _{1}\omega _{2}\omega _{3}}}\left(
-p_{0}\sum_{i=0}^{3}p_{i}q_{i}+2p_{1}p_{2}p_{3}+\mathbf{i}\sqrt{W}%
p_{0}\right)  \label{Z}
\end{equation}

Via Eq. (\ref{mu-1}, \ref{Z}, \ref{imig-1}, \ref{z1}), obtain

\begin{equation}
\mu Z=\frac{p_{0}z_{1}-p_{1}}{\left( z_{1}^{\prime }-z_{1}\right) }%
=\allowbreak \frac{1}{2}\mathbf{i}\frac{2\omega _{1}p_{1}-B_{1}p_{0}+\mathbf{%
i}\sqrt{W}p_{0}}{\sqrt{W}}.  \label{zz-1-1}
\end{equation}

Via Eqs. (\ref{mu-1}, \ref{Z}, \ref{imig-1}, \ref{rel-2}), obtain $\mu Z=%
\frac{p_{0}z_{2}-p_{2}}{\left( z_{2}^{\prime }-z_{2}\right) }$. Then, via
Eq. (\ref{imig-1}, \ref{sol-zi}), obtain

\begin{equation}
\mu Z=\frac{p_{0}z_{2}-p_{2}}{\left( z_{2}^{\prime }-z_{2}\right) }%
=\allowbreak \frac{1}{2}\mathbf{i}\frac{2\omega _{2}p_{2}-B_{2}p_{0}+\mathbf{%
i}\sqrt{W}p_{0}}{\sqrt{W}},  \label{zz-2-1}
\end{equation}

Similarly, obtain $\allowbreak \mu Z=\frac{p_{0}z_{3}-p_{3}}{\left(
z_{3}^{\prime }-z_{3}\right) }$. Then, via Eq. (\ref{imig-1}, \ref{sol-zi}),
obtain
\begin{equation}
\mu Z=\frac{p_{0}z_{3}-p_{3}}{\left( z_{3}^{\prime }-z_{3}\right) }%
=\allowbreak \frac{1}{2}\mathbf{i}\frac{2\omega _{3}p_{3}-B_{3}p_{0}+\mathbf{%
i}\sqrt{W}p_{0}}{\sqrt{W}},  \label{zz-3-1}
\end{equation}

Next we derive the stabilization equations.

(1). verify Eq. (\ref{m-1}) via $\mu Z=\frac{p_{0}z_{1}-p_{1}}{\left(
z_{1}^{\prime }-z_{1}\right) }.$

(2). verify Eq. (\ref{m-2}) via Eq. (\ref{zz-1-1}), and $z_{1}$ in Eq. (\ref%
{sol-zi}).

(3). verify: Eq. (\ref{m-3}) via Eq. (\ref{zz-2-1}),$\ $and $z_{2}$ in Eq. (%
\ref{sol-zi}).

(4). verify: Eq. (\ref{m-4}) via Eq. (\ref{zz-3-1}), and $z_{3}$ in Eq. (\ref%
{sol-zi}).

(5). verify Eq. (\ref{m-5}) via Eq. (\ref{zz-1-1}), and $z_{1},z_{2},z_{3}$
in Eq. (\ref{sol-zi}).

(6). verify \ Eq. (\ref{m-6}) via Eq. (\ref{zz-1-1}), and $z_{2},z_{3}$ in
Eq. (\ref{sol-zi}).

(7). verify Eq. (\ref{m-7}) via Eq. (\ref{zz-2-1}), and $z_{1},z_{3}$ in Eq.
(\ref{sol-zi}).

$\allowbreak $(8)$.$verify Eq. (\ref{m-8}) via Eq. (\ref{zz-3-1}), and $%
z_{1},z_{2}$ in Eq. (\ref{sol-zi}). Q.E.D.


\begin{thebibliography}{99}
\bibitem{Ferrara-1} S. Ferrara and R. Kallosh, supersymmetry and
attractors,\ Phys. Rev. D 54, 1514 (1996), hep-th/9602136;

\bibitem{Ferrara-2} S. Ferrara and R. Kallosh, universality of
supersymmetric attractors, Phys. Rev. D 54, 1525 (1996), hep-th/9603090.

\bibitem{Ceresole} A. Ceresole, R. D'Auria, S. Ferrara, and A. Van Proeyen,
Duality transformations in supersymmetric Yang-Mills theories coupled to
supergravity, Nucl. Phys. B444, 92 (1995). hep-th/9502072.

\bibitem{Behrndt} Klaus Behrndt, Renata Kallosh, Joachim Rahmfeld, Marina
Shmakova and Wing Kai Wong, STU black holes and string triality,\ Phys. Rev.
D 54, 6293 (1996). \ \

\bibitem{Behrndt-2} K. Behrndt, G. Lopes Cardoso, B. de Wit, R. Kallosh, D. L%
\"{u}st, T. Mohaupt, Classical and quantum N=2 supersymmetric black holes,
Nucl. Phys. B 488 (1997) 236.

\bibitem{Kallosh96} R. Kallosh, M. Shmakova, and W.K. Wong, Freezing of
Moduli by $N=2$ Dyons, Phys.Rev.D 54:6284-6292,1996. hep-th/9607077

\bibitem{Levay06} P. L\'{e}vay, Stringy Black Holes and the Geometry of
Entanglement, Phys. Rev. D74, 024030 (2006). e-print arxiv/hep-th/0603136

\bibitem{Duff} M.J. Duff, String triality, black hole entropy, and Cayley's
hyperdeterminant, Phys. Rev. D76, 025017 (2007).\

\bibitem{Kallosh-2021} R. Kallosh, M-theory, black holes and cosmology,
Proc. R. Soc. A, vol. 477, issue 2245, ID. 20200786. Published: 06 Jan. 2021.

\bibitem{Duff08} L. Borsten, D. Dahanayake, M. J. Duff, H. Ebrahim, and W.
Rubens, Wrapped Branes as Qubits,\ Phys. Rev. Lett. 100, 251602 (2008).

\bibitem{Borsten} L. Borsten, D. Dahanayake, M. J. Duff1, A. Marrani, and W.
Rubens, Four-Qubit Entanglement Classification from String Theory, Phys.
Rev. Lett. 105, 100507 (2010)

\bibitem{Duff-review} L. Borsten, D. Dahanayake, M.J. Duff, H. Ebrahim, and
W. Rubens, Black holes, qubits and octonions, Phys. Rep. 471, 113 (2009).

\bibitem{Linde} Renata Kallosh and Andrei Linde, Strings, black holes, and
quantum information, Phys. Rev. D73,104033 (2006).

\bibitem{Meza} A. Meza, E. A. Le\'{o}n, and J. A. Nieto, Qubits and STU
Black Holes with eight magnetic and eight Electric Charges. Arxiv:1612.02480
[gr-qc].

\bibitem{Dur} W. D\"{u}r, G. Vidal, and J. I. Cirac, Three qubits can be
entangled in two inequivalent ways, Phys. Rev. A \textbf{62}, 062314 (2000).

\bibitem{Kraus-prl} B. Kraus, Local Unitary Equivalence of Multipartite Pure
States, Phys. Rev. Lett. 104, 020504 (2010).

\bibitem{Li-1} Dafa Li, Maggie Cheng, Xiangrong Li, and Shuwang Li,
Classification of \textquotedblleft Large\textquotedblright\ black holes
into seven families, Quantum Information and Computation, Vol. 24, No. 7\&8
(2024) 0576-0608.

\bibitem{Li-2} Dafa Li, Maggie Cheng, Xiangrong Li, and Shuwang Li, SLOCC
and LU Classification of Black Holes with Eight Electric and Magnetic
Charges, Int J theor phys 63, issue 6, 144 (2024).

\bibitem{Ceresole-2} A. Ceresole, R. D'Auria and S. Ferrara, The Symplectic
Structure of N=2 Supergravity and its Central Extension,
Nucl.Phys.Proc.Suppl. 46 (1996) 67-74. e-print: hep-th/9509160.
\end{thebibliography}
\end{document}